\documentclass[11pt]{elsarticle}
\usepackage{graphicx}
\usepackage{color}
\usepackage{tikz}
\usetikzlibrary{arrows,shapes,scopes}
\usetikzlibrary{calc}
\usetikzlibrary{graphs}
\usepackage{hyperref}
\hypersetup{colorlinks=true,linkcolor=blue,citecolor=purple}
\usepackage{url}

\begin{document}
\begin{frontmatter}

\title{Distributed dynamic load balancing for task parallel programming\tnoteref{t1}}
\tnotetext[t1]{This work was supported in part by the Swedish Research Council and carried out within the Linnaeus centre of excellence UPMARC, Uppsala Programming for Multicore Architectures Research Center.}

\author{Afshin Zafari}
\ead{afshinzafari@gmail.com}

\author{Elisabeth Larsson\corref{c1}}
\ead{elisabeth.larsson@it.uu.se}
\cortext[c1]{Corresponding author}

\address{Uppsala University, Department of Information Technology, Box~337, SE-751~05 Uppsala, Sweden}

\begin{abstract}
In this paper, we derive and investigate approaches to dynamically load balance a distributed task parallel application software. The load balancing strategy is based on task migration. Busy processes export parts of their ready task queue to idle processes. Idle--busy pairs of processes find each other through a random search process that succeeds within a few steps with high probability. We evaluate the load balancing approach for a block Cholesky factorization implementation and observe a reduction in execution time on the order of 5\% in the selected test cases.
\end{abstract}

\end{frontmatter}

\section{Introduction}

The objectives of improving the load balance across computational resources can be to reach the best possible utilization of the resources, to improve the performance of a particular application, or to achieve fairness with respect to throughput for a collection of applications. Load-balancing can be static, decided a priori, or dynamic, that is, changing during the execution of the application(s). An overview of various issues related to dynamic load balancing (DLB) is given in~\cite{Alakeel10}. DLB strategies do not assume any specific pre-knowledge about the application. However, the strategies are still often based on some assumptions on the type of parallelization or the class of algorithms.


DLB can be implemented through data migration~\cite{Balasubramaniam04,Martin13}. This is especially relevant when the algorithm consist of iterations or time steps, where similar computations are repeated. It is then likely that a redistribution of data will be effective over at least a few consecutive iterations. 

If the parallel implementation is instead task centric, where a task is a work unit, another possibility is to migrate computational work~\cite{Khan10,Rubensson14}. As work is usually associated with data, this could also include moving data, temporarily or permanently.

An approach, that currently is receiving attention, instead migrates the computational resources between jobs or between processes~\cite{Spiegel06,Garcia12,Schreiber15,Garcia17}. The underlying assumption is that there is a hybrid parallelization, where MPI is used over the computational nodes in a cluster, but shared memory, thread-based parallelization is used within the computational nodes. Assuming that several processes are co-scheduled within one computational node, resources can, based on the malleability of the shared memory tasks,  be migrated within the node. This approach alone cannot provide global load balance, but can improve utilization within the node, and can adjust global imbalance depending on the mix of processes at the node. 
%






In this paper, we consider distributed DLB in the context of distributed task-based parallel programming~\cite{TFGBAL11,AAFNT12,BBDFHD13,Rubensson14,Zafari17} with a hybrid MPI-thread implementation. We are not assuming co-scheduling of several processes, instead we are aiming to improve the performance of a single application running on a cluster of multicore nodes.

Task stealing as mechanism for load balancing has been proven efficient in the shared memory task-based parallel programming context, e.g., in the Cilk~\cite{cilkplus} C++ language extension and the SuperGlue~\cite{Tillenius15} framework. It is also used in the distributed task framework Chunks and Tasks~\cite{Rubensson14}. 

This is the direction that we are investigating also in this paper. We derive a prediction model to decide if stealing is likely to improve utilization, in the sense that the work can be finished by the remote process, and the result returned earlier than it was processed in the current location. All decisions are taken locally to avoid bottlenecks due to global information exchange or centralized scheduling decisions. The resulting load balancing approach is implemented within the DuctTeip distributed task parallel framework~\cite{Zafari17}.

The paper is organized as follows: In Section~\ref{sec:task} we briefly review he aspects of task parallel programming that are important for DLB. The properties of the DLB approach we are proposing are described in Section~\ref{sec:methods}. Then a theoretical analysis for when it is cost efficient to export tasks is performed in Section~\ref{sec:theor}. The Cholesky benchmark used for evaluating the method is described in Section~\ref{sec:chol}, while the results of the performance experiments are given in Section~\ref{sec:exp}. Finally, conclusions are given in Section~\ref{sec:conc}.

\section{Definition of the task parallel programming context}\label{sec:task}
In our load balancing model, we do not make any assumptions about the application as such, but we target dependency-aware task parallel implementations. We further assume that the distributed application is executed by a number of (MPI) processes $p_i$, $i=0,\ldots,P$ and that each process has a queue of ready tasks to execute. Tasks become ready when their data dependencies are fulfilled and the data they need in order to run are available locally. 

In the DuctTeip framework~\cite{Zafari17}, where we will implement the DLB strategy, the default situation is that a certain task is executed by the process that owns the output data of the task. That is, the data distribution also determines the task distribution. For a data parallel algorithm, this may be sufficient to achieve a reasonable load balance by a uniform splitting of the data. However, if some of the processes are slowed down due to, e.g.,  external interference, there can still be imbalance in the end. For more complex algorithms, it is expected that the work load of the individual processes will vary over the execution.  

A run-time system handles all task management decisions, such as checking when tasks are ready to run, and sending and receiving data from remote processes. The run-time system also handles DLB. We consider the possibility that the run-time system records performance data for different task types, and for the communication, but we do not assume that this requires modification of the user code or of the operating system.

\section{The dynamic load balancing approaches}\label{sec:methods}
We start by defining the workload $w_i(t)$ of process $p_i$ at time $t$ as the number of ready tasks in the queue. This does not take the size of the tasks into account, but it is an easily accessible number that can be stored as one integer variable per process. What is a high (or low) workload depends on the application, the blocking of the data, and the number of processes $P$. We let the threshold $W_T$ be a user defined parameter, and then define processes with $w_i>W_T$ as busy and processes with $w_i\leq W_T$ as idle. A more correct definition is to say that a process is idle when $w_i=0$, but in this case, we want the processes to start looking for more work before they run out of it. In this way, the migration of tasks can overlap with computational work.

\begin{figure}[!htb]
\centering
\includegraphics[width=0.49\textwidth]{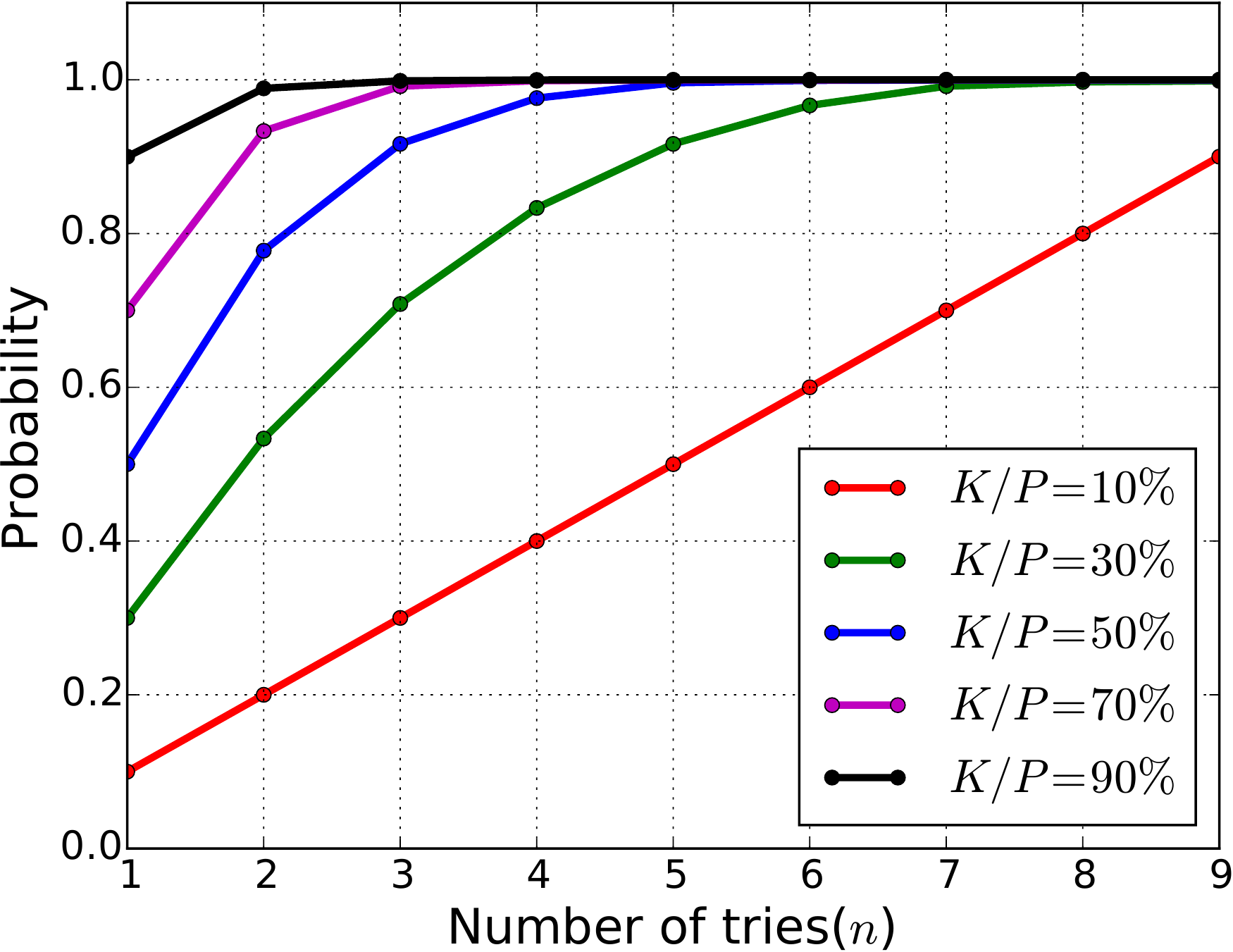}
\includegraphics[width=0.49\textwidth]{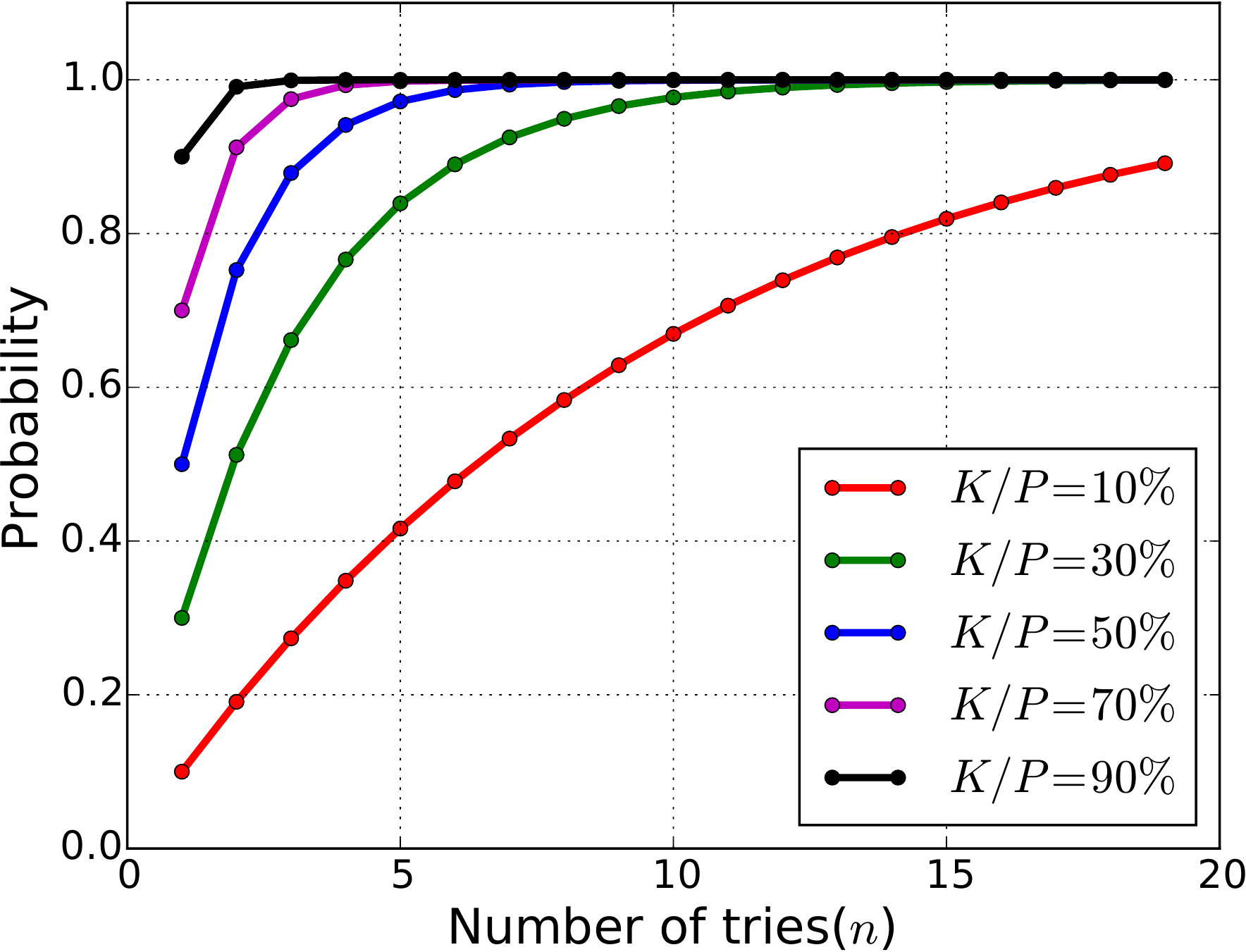}
\caption{The probability of success for finding any of the $K$ busy processes out of a total of $P$ using $n$ tries for $P=10$ (left) and $P=100$ (right).}\label{fig:prob}
\end{figure}

Obtaining global information about the workload of all processes is likely to become a bottleneck when scaling to larger numbers of processes, and we want to avoid this and let each process make local decisions. The idea that we are using is that each process periodically tries to become a partner in an idle--busy process pair. We do not consider any particular topology of the network, but let the processes randomly try other processes with a uniform selection probability.  The probability $\mathcal{P}(k)$ of finding $k$ busy processes in $n$ tries drawn from a distribution where $K$ processes out of a total of $P$ are busy, is given by the hypergeometric probability distribution
\begin{equation}
\mathcal{P}(k) = \frac{
\left(\begin{array}{c}P-K\\n-k\end{array}\right)
\left(\begin{array}{c}K\\k\end{array}\right)
}{
\left(\begin{array}{c}P\\n\end{array}\right)
}.
\end{equation} 
The probability of at least one successful try out of $n$ is the complementary probability of failure, that is, $1-\mathcal{P}(0)$. This function is plotted for different combinations of $P$ and $K$ in Figure~\ref{fig:prob}. 
Since both idle and busy processes are looking for each other, the most difficult case is when 50\% of the processes are idle/busy. By analyzing the formula, we find that for $K=P/2$, as the number of processes $P\rightarrow\infty$, the probability of success approaches $1-2^{-n}$, that is for $n=5$ tries, the probability is more than 96\%. We therefore decide that a process looking for a partner will always perform 5 tries, then wait for a period $\delta$ before starting another round of tries. This waiting time is introduced to prevent flooding the network with requests when there is no work to share. The waiting time is the second user defined parameter that needs tuning. A successful request means that the pair of nodes will not accept or send any further requests until their work exchange transaction has completed.


When a busy--idle process couple has been formed, the next step is to decide which tasks to export, if any. We consider three potential strategies
\begin{enumerate}
\item \textbf{Basic:} No extra information is exchanged. The busy process $p_i$ just sends its excess tasks such that the remaining queue is $w_i=W_T$.
\item \textbf{Equalizing:} The idle process $p_j$ appends information about its current work load $w_j$ to the request. The busy process $p_i$ computes the average $\bar{w}=(w_i+w_j)/2$, and sends $w_i-\bar{w}$ tasks to $p_j$. 
\item \textbf{Smart:} The idle process provides information about the expected time to execute the currently enqueued tasks. The busy process estimates which of the tasks would return their results earlier if executed remotely, than the time they would be completed if executed locally. Only the tasks with an expected benefit are exported.
\end{enumerate}
In the latter case, performance estimates are needed. Each process records the average time for running tasks of each type as well as times for communicating task of each type and data of a certain size. The sophistication of the models applied to the measurements can vary, but they will be used in the same way. The cost for remote execution consists of  the remote queuing time, the time for exporting the task and its data, the task execution time, and the communication time for returning the result, while the time for local execution is the local queuing time and the task execution time.

In the method described above, there is only one threshold parameter, and all tasks are either busy or idle. An alternative would be to have a gap between the idle and busy levels. This would reduce the number of requests as some processors would be in the middle zone. Also, it could reduce the risk for overshooting in the sense that a processor that was idle, but close to the threshold immediately becomes busy after receiving work from its busy partner. 

\section{A theoretical analysis of the cost for task migration}\label{sec:theor}
With more knowledge about the tasks and the hardware we can perform better predictions for which tasks to share and how many to export when a work request arrives. However, this also makes the approach more intrusive in the sense that the application programmer needs to provide more information. Here we will look at the cost in time for executing a task remotely compared with executing it in location.

Assume that a computational node in the considered hardware performs $S$ floating point operations per second, and can deliver $R$ doubles per second from the main memory. When exporting a task, we need to send the input data together with the task, and then we need to return the output data. Let the total number of doubles in the input and output data be $D$, and let the number of floating point operations performed by the task be $F$. Then the time for executing the task locally is
\begin{equation}
T_L = F/S,
\end{equation}
and the time for executing the task remotely and returning the data is
\begin{equation}
T_R = F/S + D/R.
\end{equation}
The fraction of extra time that is needed for remote execution is given by
\begin{equation}
Q = \frac{S}{R}\frac{D}{F}.
\end{equation}
For a modern computer system, floating point operations are faster than data transfer, and a typical ratio can be around 40. This is the case for the system used for the experiments in Section~\ref{sec:exp} (see~\cite{Zafari17} for a detailed calculation). The second ratio $D/F$ represents the computational intensity of the task. 

If we, e.g., consider a block matrix--matrix multiplication, with blocks of size $m\times m$, then $F=2m^3$, and $D=3m^2$. This leads to a total ratio of $Q = 60/m$. That is, for such a task, the cost for remote execution is almost negligible if the block size is large enough. 

If we instead consider a matrix--vector multiplication task, the situation is different. Then $F=2m^2$ and $D=m^2$ leading to $Q=20$. That is, 20 tasks can be executed locally in the same time as one task is migrated, executed remotely, and the result returned.

By looking at these numbers we can get an understanding for how the threshold parameter $W_T$ should be chosen. For computationally intensive applications, a rather small value will be sufficient to make sure that there is local work to cover the cost for exporting tasks. However, if the tasks are less computationally intensive, it is not worth exporting tasks until the local work load is very high with, in this case, more than 20 tasks left in the queue for each exported task.

\section{The Cholesky benchmark}\label{sec:chol}
\begin{figure}[!htb]
\centering
\raisebox{1cm}{\includegraphics[width=0.5\textwidth]{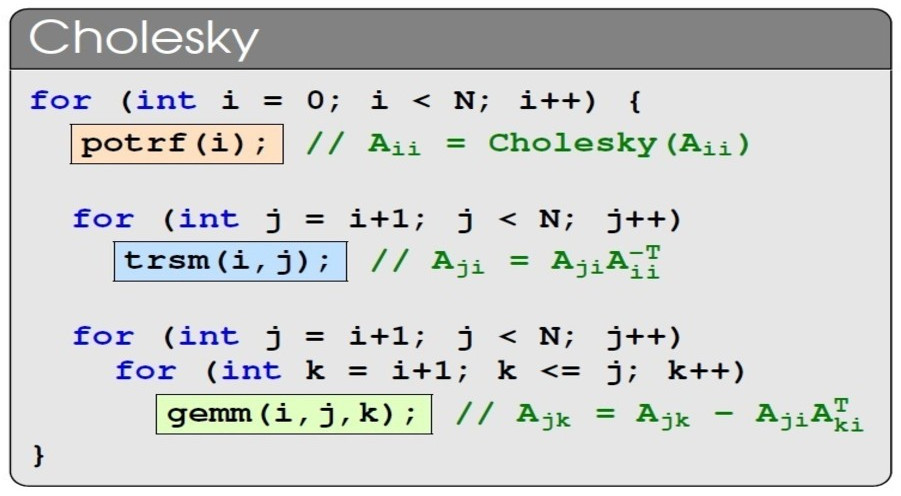}}\hspace{1cm}
\includegraphics[width=0.25\textwidth]{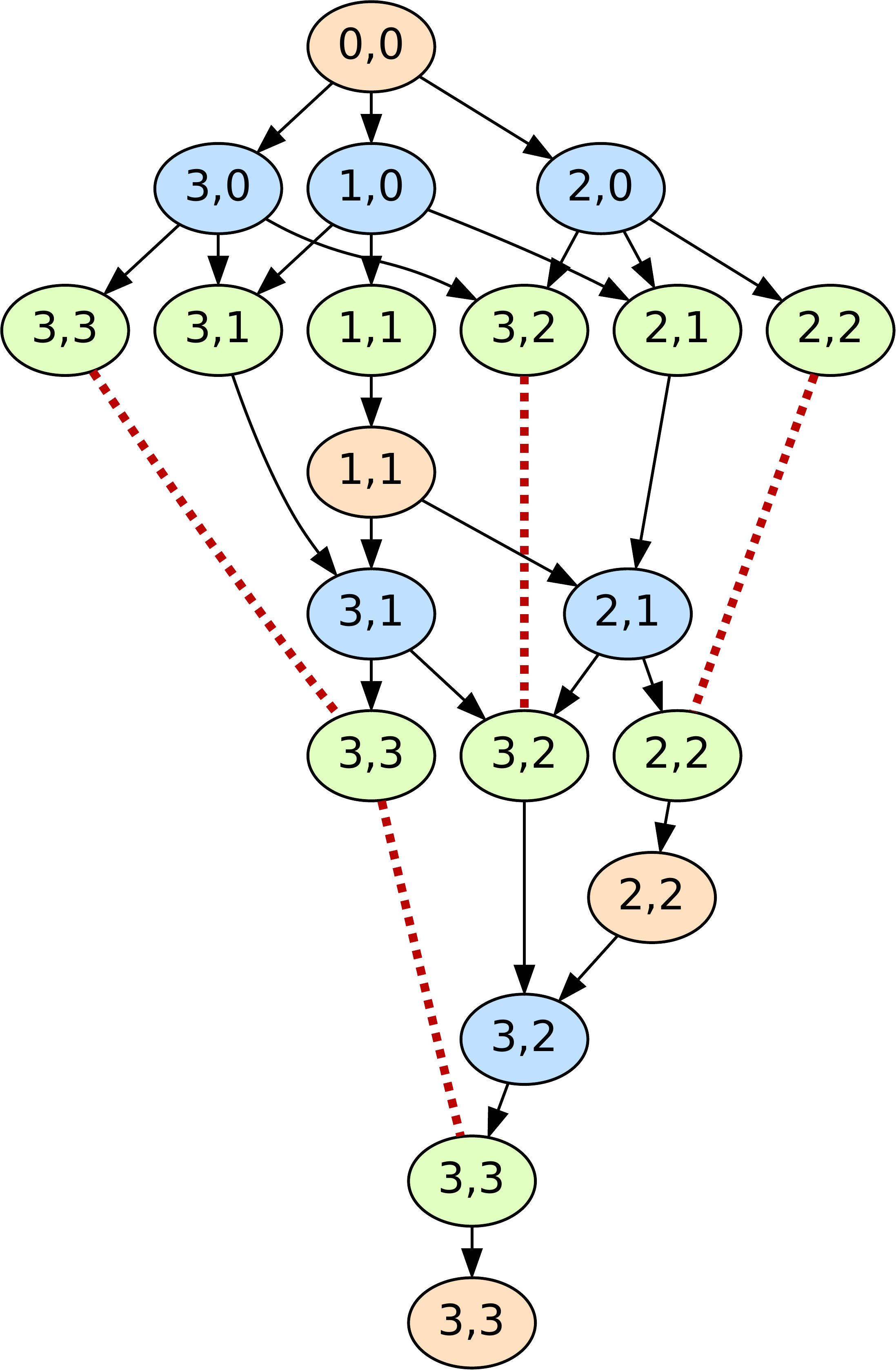}
\caption{The Cholesky algorithm (left) and a Cholesky task graph for a $4\times4$ block matrix (right). In the algorithm, the subroutine calls in colored boxes are implemented as tasks. $N$ is here the number of blocks. The numbers in the task graph correspond to the indices of the block that is updated, the solid lines indicate must-execute-before dependencies, while the dashed lines correspond to tasks that can be performed in any order, but not at the same time.}\label{fig:chol}
\end{figure}

We use a right-looking block Cholesky factorization as a benchmark problem to investigate the performance of the suggested DLB mechanisms. Most of the tasks in this application are computationally intensive, which makes it a good candidate for success. The algorithm is implemented with the DuctTeip framework and DLB can be turned on or off. The algorithm starts from the leftmost column, first the block on the main diagonal is factorized, and then the blocks below the diagonal are updated. Finally the blocks to the right of the column are updated. This procedure continues until all columns have been factorized. Since the input matrix is symmetric, only the lower triangular part is used in the algorithm. During the execution of the algorithm there is a data flow from the top rows and first columns to the bottom rows and last columns of the matrix. The algorithm and its corresponding task graph are illustrated in Figure~\ref{fig:chol}.

The matrix blocks are distributed block cyclically onto the virtual process grid. The amount of communication as well as the load imbalance (see e.g.,~\cite{scalapack,Zafari17}) is minimized when the process grid is square. This is not always possible, and here we instead consider cases where the number of processes is a prime number or a product of two different prime numbers. The non-square configurations lead to significant load imbalance, and we investigate if DLB can improve the performance in these cases.

\section{Experimental results}\label{sec:exp}

The performance experiments have been performed at the Rackham cluster at Uppsala Multidisciplinary Center for Advanced Computational Science (UPPMAX), Uppsala University. The cluster currently has 334 dual socket nodes with 128 GB/256 GB memory each. Each socket is equipped with a 10 core Intel Xeon E5 2630 v4 (Broadwell) processor running at 2.2 GHz. When running distributed applications at the cluster, a number of complete computational nodes are allocated. That is, no other application codes are running at the same nodes. The experiments are performed on applications running within the DuctTeip task parallel framework.

In order to run an application with DLB, we need to find appropriate values for the work load threshold $W_T$, and the waiting time $\delta$. A suitable threshold value should depend on the application work load over the execution time. For the experiments performed here, it is determined offline by first running the application once without DLB, and then setting $W_T=\max_{i,t}w_i(t)/2$. For a production DLB version, the threshold could for example be initialized with a reasonable starting value, and updated locally by each process in relation to the local work load. In the basic model, selecting $W_T$ as described above corresponds to a behavior that resembles that of the equalizing model, as approximately half of the tasks will be exported for a busy process. 
 
\begin{figure}[!htb]
\centering
\includegraphics[width=0.5\textwidth, height=.25\textheight]{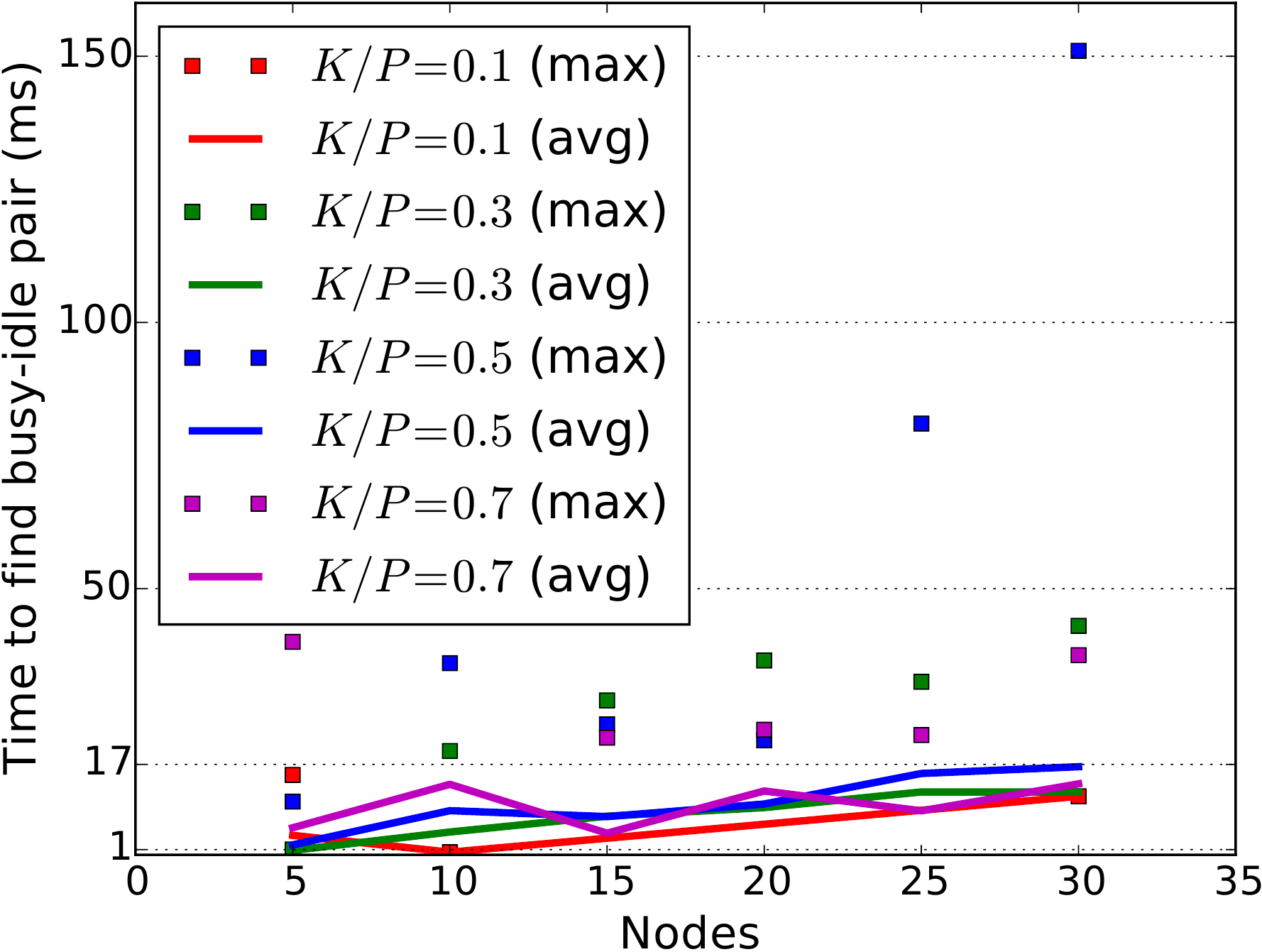}
\caption{The average time for finding a busy--idle process pair.}\label{fig:delta}
\end{figure}

The waiting time $\delta$ should instead depend on the network bandwidth and should be long enough to allow the waiting process to be found by a partner. We performed several experiments to find the expected time required for finding a busy--idle process pair. The experimental results are shown in Figure~\ref{fig:delta}. Both the average times and the maximum times are plotted. As expected, the average time grows slowly with the number of processes, and is largest for equal fractions of busy and idle processes.
%

\begin{figure}[!htb]
\centering
\includegraphics[width=0.49\textwidth]{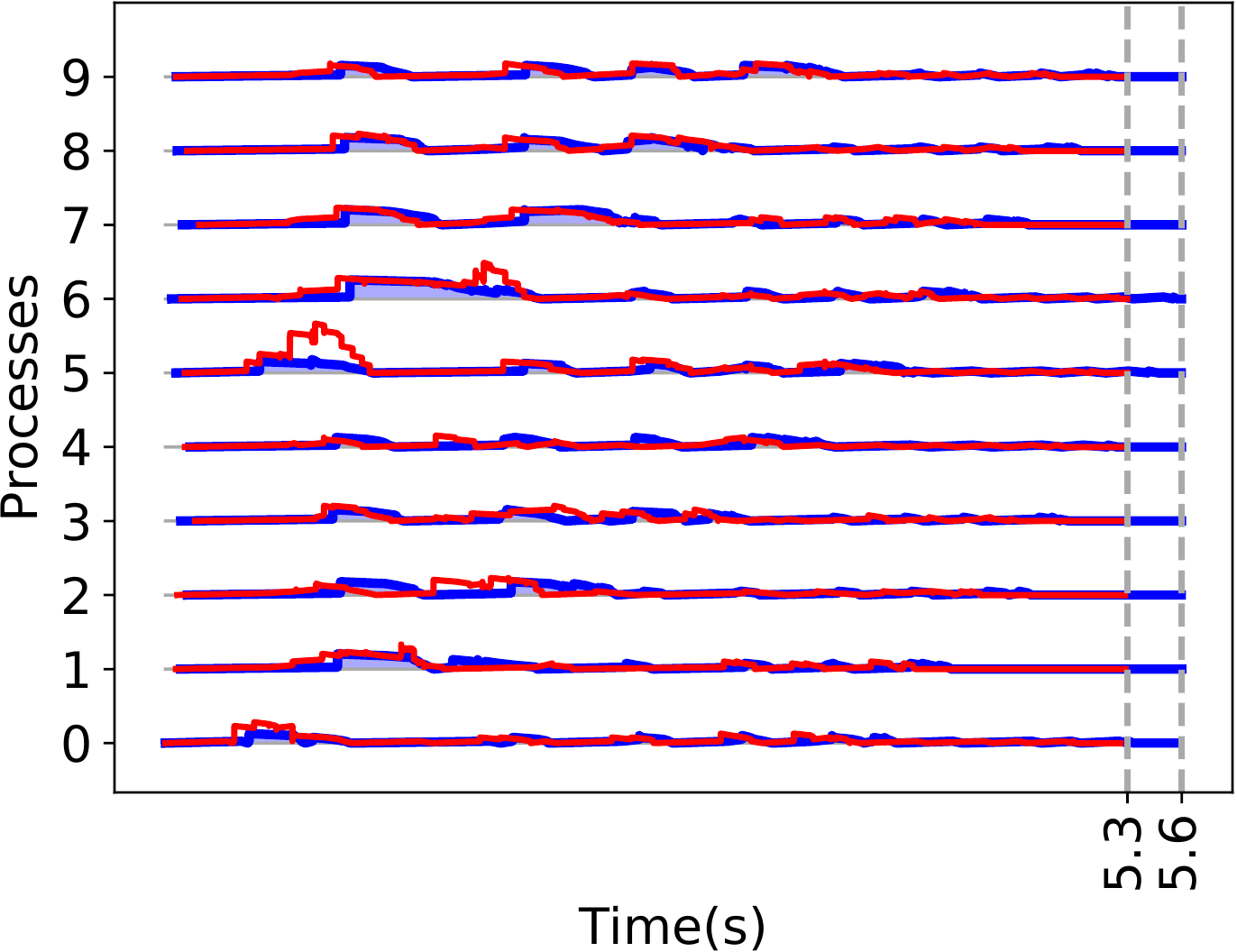}
\includegraphics[width=0.5\textwidth]{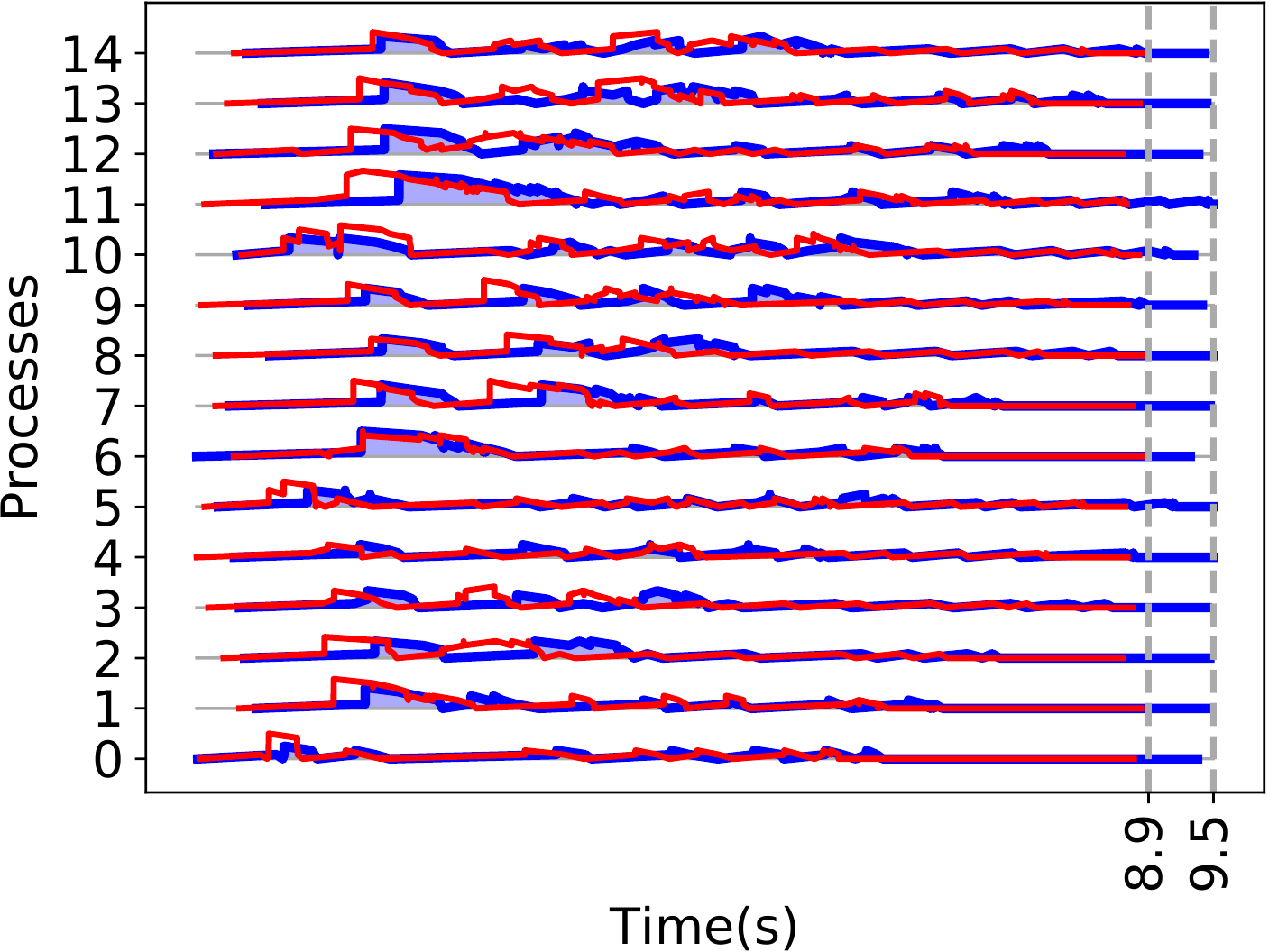}
\caption{The work load for each process in the Cholesky factorization for matrix size $N=20\,000$, and $P=10$ processes arranged in a $2\times5$ process grid (left), and for matrix size $N=30\,000$, and $P=15$ processes arranged in a $3\times5$ process grid (right) without DLB (filled blue curves) and with DLB (red curves). }\label{fig:dlbsxsful}
\end{figure}

In the following experiments, 10--15 processes are used, and according to the results in Figure~\ref{fig:delta}, a waiting time of $\delta=10$ ms is a suitable value. The maximum work load over the execution for any process is $w_i=10$, and the threshold is chosen as $W_T=5$. 
Figure~\ref{fig:dlbsxsful} shows the workload and execution times for the Cholesky factorization for two different problem sizes and process grids. In both cases, the matrices are divided into $12\times12$ blocks and distributed block-cyclically over the processes. Here, the application of DLB is successful in both cases, and the total execution time is reduced by 5--6\%. In some places, one can see that one process is much more loaded with DLB than without. These can be cases where an equalizing approach would be more beneficial.

\begin{figure}[!htb]
\centering
\includegraphics[width=0.5\textwidth]{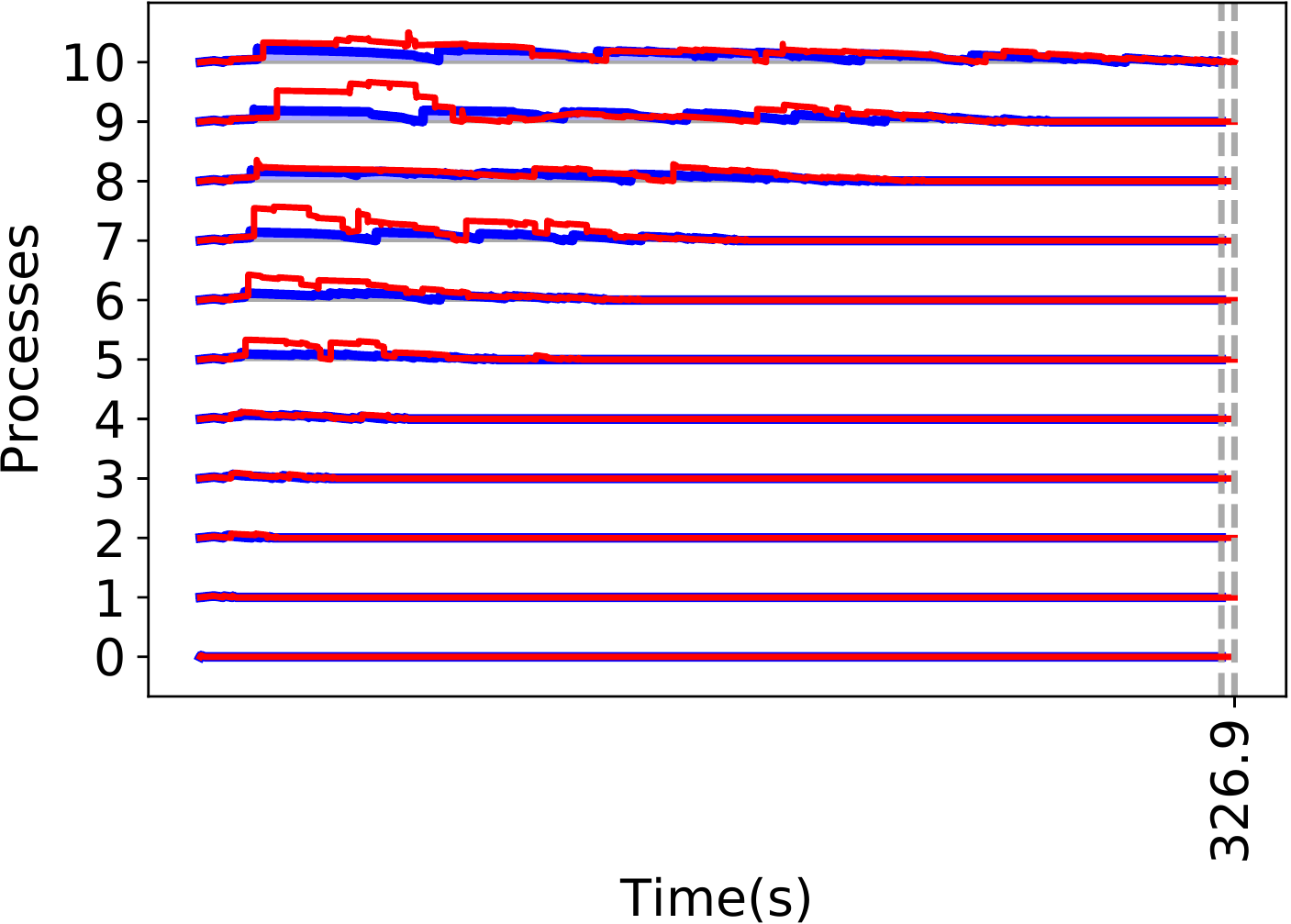}%
\includegraphics[width=0.5\textwidth]{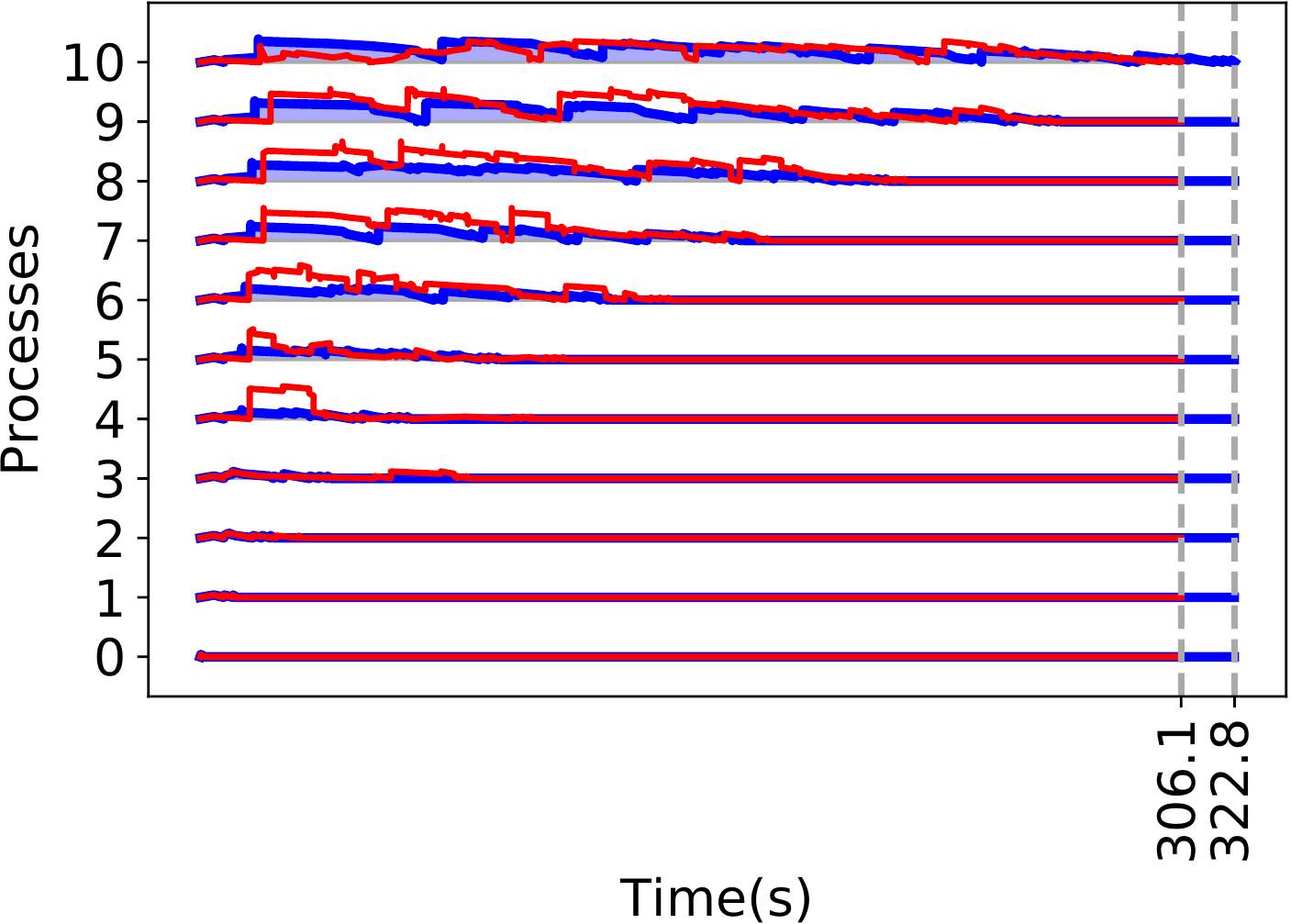}
\caption{The work load for each process in the Cholesky factorization for matrix size $N=100\,000$, and $P=11$ processes arranged in an $11\times 1$ process grid, without DLB (filled blue curves) and with DLB (red curves). Two executions are shown, one unsuccessful (left) and one successful (right).}\label{fig:dlbfail}
\end{figure}

The process of randomly selecting partners for work migration and the variability of work load and type of tasks between different processes within an application make the results of applying DLB non-deterministic. In Figure~\ref{fig:dlbfail}, two executions of the same application configuration are shown, where one is successful, while the other one fails to provide any improvement. The matrix is here divided into $11\times 11$ blocks, which matches the number of processes.
\section{Conclusions}\label{sec:conc}
We have discussed how to create a low overhead DLB functionality in a task parallel programming framework. An important aspect of the approach is that all decisions are local and the processes act autonomously, hence avoiding bottlenecks due to global exchange of information.

Processes that either have a high or low work load search for another process with the opposite load situation to share work with. This search is randomized. This could be a disadvantage if the communication is much more expensive when the computational nodes are far from each other. Then processes could be grouped and DLB be applied within the group. However, an advantage compared with for example diffusion-based DLB~\cite{Khan10} is that load can be propagated to anywhere in the system, while diffusion needs to go via nearest neighbors.

Very few assumptions are made in the model apart from the assumption that the context is a distributed task parallel run-time system. However, the threshold parameter $W_T$ is application dependent, at least in the basic model. The theoretical analysis in Section~\ref{sec:theor} can be used as a guideline for deciding on $W_T$. The waiting time $\delta$ can be determined once for a particular system.

In the Cholesky experiments that we have performed, even though the load imbalance was not extreme, we could see that the basic DLB version could give improved performance. It is therefore of interest to perform further experiments and to develop the DLB model further.

\section*{Acknowledgments}
The computations were performed on resources provided by SNIC through Uppsala Multidisciplinary Center for Advanced Computational Science (UPPMAX) under Project SNIC 2017/1-448.

\bibliographystyle{siam}

\end{document}